# Situation-Aware Approach to Improve Context-based Recommender System

Djallel Bouneffouf
Télécom SudParis
9, rue Charles Fourier
91011 Evry, France

Djallel.Bouneffouf@it-sudparis.eu

## ABSTRACT
In this paper, we introduce a novel situation-aware approach to improve a context based recommender system. To build situation-aware user profiles, we rely on evidence issued from retrieval situations. A retrieval situation refers to the social-spatiotemporal context of the user when he interacts with the recommender system. A situation is represented as a combination of social-spatiotemporal concepts inferred from ontological knowledge given social group, location and time information. User's interests are inferred from past user's interaction with the recommender system related to the identified situations. They are represented using concepts issued from a domain ontology. We also propose a method to dynamically adapt the system to the user's interest's evolution.

## Categories and Subject Descriptors
H.3.3 [**Information Search and Retrieval**]: *Information filtering, Selection process, Relevance feedback.*

## General Terms
Algorithms

## Keywords
Personalization, context, contextual preferences, user profile modeling, personalized access.

## 1. INTRODUCTION
Mobile technologies have made access to a huge collection of information, anywhere and anytime. This brings big challenges for the Recommender System (RS) field. Indeed, technical features of mobile devices yield to navigation practices which are more difficult than the traditional navigation task.
Studies on logs of mobile Internet user interactions [1] show that most of the information needs of mobile users are related to contextual factors such as the user's interests, his social group, the location and the time of the interaction.
Recent works in RS community attempt to improve accuracy in this field. These works aim at filtering large amounts of information and return a view on the information which matches the user's preferences and interests. Contextual Recommender System (CRS) process is a key concern in these works, including the need to provide information tailored to an individual user and taking the context into account. While some works use only the user feedback to build the user profile ([2], [3]), other ones ([4], [5]) use contextual information, issued from his external environment, as an additional source of evidence to build dynamic user profiles.
In order to give CRS systems the capability to provide a mobile user with information matching his interests adapted to his situation, our contribution consists in abstracting from sensor data some semantic information to characterize situations in which a user interacts with the CRS. A user preference is learnt for each identified situation, on the basis of past interaction activities occurred in this situation. A case base reasoning method is then applied to dynamically select the most appropriate recommendation by comparing the current situation with the already learnt ones.
The paper is organized as follows. Section 2 reviews some related works. Section 3 presents our approach for representing and building a situation-aware user profile for a mobile user of CRS. Section 4 presents our method for exploiting the user profile in a contextual recommender system. The last section presents our conclusion and points out possible directions for future work.

## 2. Related work
A considerable amount of research has been done in recommending relevant information for mobile users. Earlier techniques ([2], [3]) are based solely on the computational behavior of the user to model his interests regardless of his surrounding environment (location, time, near people). The main limitation of such approaches is that they do not take into account the dynamicity of user interests regarding his situation.
Few research works attempted to tackle this limitation. In [4] the authors propose a method which consists of building a dynamic profile based on time and user experience, where the user preferences and interests are weighted according to temporal zones. These zones are learnt by studying the user's activities during different periods of time. To model the change of user's preferences according to his temporal situation in such periods (workday, vacation), the weighted association for the concepts in the user profile is established for every new experience of the user.
In [5] they propose a context-aware mobile RS for young people in leisure time. The system predicts the user's current and future leisure activity (eating, reading and shopping) from context (time, location) and user behavior. The predicted user activity combined with models of the user's preferences, are used together to filter and recommend relevant content.
The work in [6] describes a method for generating metadata for photos using spatial, temporal and social information. They describe a system that allows inferring location information to photos taken with a phone. In particular, the combination and sharing spatial, temporal, social and contextual metadata from a particular user and between users allows drawing conclusions about media content.

Another work [10] describes a RS operating on three dimensions of context that complement each other to get highly targeted. First, it analyzes information such as clients' address books to estimate the level of social affinity among users. Second, it combines social affinity with the spatiotemporal dimension and the user's history in order to improve the quality of the recommendations.

In [11], the authors present a technique to perform user-based collaborative filtering (CF). Each user's mobile device stores all explicit ratings made by its owner as well as ratings received from other users. Only users in spatiotemporal proximity are able to exchange ratings and they show how this provides a natural filtering based on social contexts.

In our approach, we exploit both the history of the user-system interaction and diverse ontologies (location, time and near people) to learn user's situations and their corresponding user's interests. In comparison with the existing works, we have added the following new features:

- A rich semantic representation of the user interaction situations as concepts from social, location and time ontologies with their corresponding user's interests, while in [10,11] the user situation is represented by low level data.
- Automatic process of building user's profile: no effort is needed from the user, while in [4] the user is solicited.
- No restriction on user's situations or population, while in [5, 6] it is devoted to some specific situations and specific populations.
- Three dimensions of the context are considered (social, spatial and temporal), while in [4, 5] they only consider two.

In what follows, we define the structure of the proposed user model and the methods for inferring the recommendation situations from low level information (Section 3). Then, we explain how to build dynamic user profiles, and how dynamically select the adequate user's preferences, according to the current situation, in order to recommend relevant information (Section 4).

## 3. Defining the user's model

In mobile CRS, the user's profile may change anytime due to changes in the user's environment (location, time, near persons, etc). As stated in Section 2, static approaches for building the user's profile ([2], [3]) are therefore poorly useful, so we rather focus on more dynamic techniques, capable of continuously adjusting the user's interests to the current situation. To distinguish mobile user's situations when the system recommends information, we propose a situation to be composed of location, time and social characteristics. We also propose a dynamic user's profile that can be used to recommend relevant information according to user's information needs in a certain situation.

A user U is represented by a set of situations with their corresponding user's preference, denoted: U= (S, UP), where S is a situation and UP its corresponding user's preference.

Figure 1 depicts the entity-relationship (ER) diagram of the proposed user model (by the sake of simplicity, we omit entities' attributes).

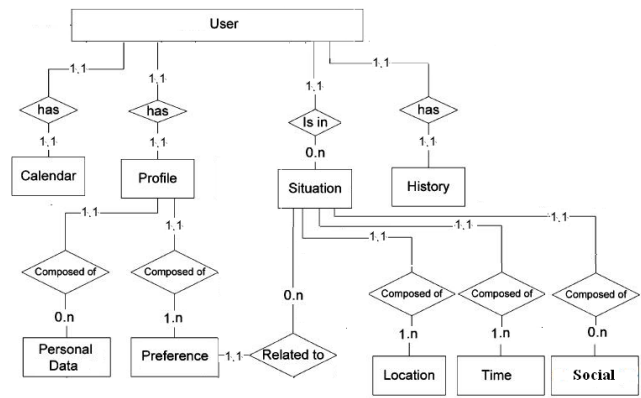

**Figure 1 : ER diagram of the user model**

Looking at Figure 1, the user's profile is modelled by an entity related to a Personal Data entity and a Preference entity. Preference, Situation and History entities are detailed below.

### 3.1 Preference

The preference of a user is a complex notion. Let us take an example. Paul prefers French food when he goes to the restaurant in France, as usual. When he travels, he prefers local food. The condition (when he travels or when he is in France) is related to the situation of the individual. Paul's preference concerning food in each situation is what we refer to as Preference. As we can see from this example, preferences are contextual, and might depend on many factors that range from one's own location to a friend's situation. In the user's model (Figure 1) we relate Preference and Situation entities through a *Related to* relationship. Preferences are built after navigation activities done by the user; they contain the set of documents of the navigation made by the user in the current situation. A navigation activity expresses the following sequence of events: the user opens the system and navigates to get the desired information; the user expresses his preferences on the documents. We assume that a document is relevant if there are some observable user's behaviours, like clicking or saving it.

### 3.2 Situation

Our challenge when building a dynamic profile is to use sensory data to identify a user's situation. We propose to associate low level information directly acquired from sensors to semantic concepts extracted from social, temporal and spatial ontologies.

Hence, suppose the user is associated to: the location "38.868143, 2.3484122" from his phone's GPS; the time "Mon Oct 3 12:10:00 2011" from his phone's watch; and the meeting with Paul Gerard from his agenda. From this knowledge, we infer that he is "in a restaurant, with the general manager of the wine French company "Le Chateau Fort" at midday, and it is a workday".

Thus, we consider a situation composed of the following three dimensions.

**Location**

There are different ways to characterize a location. As returned by location sensor systems (like GPS), location is a position in systems based on geographic coordinates, or may also be defined by an address. Simple automated place labelling systems are already commercialized (Google map, Yahoo local...) and consist of merging data such as postal addresses with maps.

In our user model, we use a spatial data base and a spatial ontology to represent and reason on geographic information. We propose a mapping between the geographic coordinates in the spatial data base and semantic representations in the spatial ontology by their name and type (e.g. "The magnum" restaurant).

**Time**
The temporal information is complex: it is continuous and can be represented at different levels of granularity. To define the temporal aspects characterizing the user's situation, we suggest abstracting the continuum time into some specific and significant periods (abstract time classes), which we expect having an effect on the user behavior (e.g. morning, weekend). To allow a good representation of the temporal information and its manipulation, we propose to use OWL-Time ontology [7] which is today a reference for representing and reasoning about time. We propose to base our work on this ontology and extend it if necessary.

**Social**
The social dimension refers to the information of the user's interlocutors (e. g. the user is with his friend, with an important client, with a colleague or with his manager...).
To define the near people aspects characterizing the user's, a clear model for the representation and reasoning on social clustering is necessary. We use the FOAF Ontology [12] to describe the social network by a set of concepts and properties.

In Figure 1, a situation and each of the corresponding dimensions are represented by an entity (Situation, Location, Time and Social, resp.). More specifically, a situation S can be represented as a triple whose features X are the values assigned to each dimension: $S = (X_l, X_u, X_v)$ where $X_l$ (resp. $X_u$ and $X_v$) is the value of the location type (resp. time and social) dimension.

## 3.3 History
Figure 3 shows the ER diagram of the history model. The history is composed of a Presence entity and an Information entity: the former represents the triple of entities (Location, Time, Social) i.e. a Situation (as explained in Section 3.2); the later is composed of all requested and received data information that are spread between the user and the system.

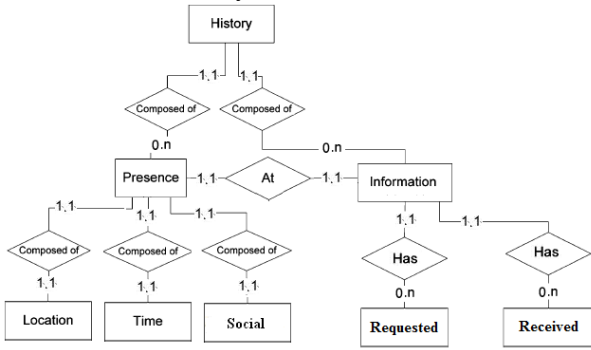

**Figure 2 : ER diagram of the history model**

## 4. Updating profile and recommend items

In order to select the most adequate user's preference to be used for recommendation in a given situation, we use case-based reasoning (CBR) technique to compare the similarity between the current situation and the past ones. In CBR, a problem is solved based on similar solutions of past problems [8]. A case is usually described by a pair (premise, value). Premise is the description of the case containing its characteristics, while value is the result of the reasoning based on the premise. A previous experience, which has been captured and learned, is referred to as a past case.

Likewise, a new case is the description of a new problem to be solved.
In our situation-aware computing approach, the premise part of a case is a specific situation S of a mobile user when he navigates on his mobile device, while the value part of a case is the user's preference UP to be used for the recommendation. Each case from the case base is denoted as C= (S, UP), being C the relationship between S and UP in our user model (Figure 1).
Our CBR approach involves the following four steps.

**Step 1: Identifying the Current Situation**
To represent the current situation S, sensory data are abstracted from the time, location and social ontologies using GPS sensor, system's clock and user's agenda data, as outlined in section 3.2. We obtain then a semantic representation of S, $(X_l, X_u, X_v)$, being $X_l$, $X_u$ and $X_v$ the semantic representations of location, time and social dimensions, resp.

**Step 2: Retrieving the Most Similar Situation**
To determine the expected user preference in the current case C, the current situation S is compared to the past ones. Let $PS = \{S^1,....,S^n\}$ be the set of past situations. Then, we select from PS the situation $S^p$ using the same similarity measure as [10, 15, 16, 17, 18, 19] defined by:

$$S^p = \arg\max_{S^i \in PS} \left( \sum_j sim_j(X_j, Y_j^i) \right) \quad (4)$$

Where $X_j$ is the value of the j dimension of the situation vector S, $Y_j^i$ is the value of the j dimension of the situation vector $S^i$ existing in the case base PS, $sim_j$ is the similarity metric related to the j dimension between two situation vectors.
The similarity between two concepts of a dimension j depends on how closely they are related in the corresponding ontology (location, time or social).
We use the same similarity measure as [10] defined by:

$$sim_j(X_j, Y_j) = 2 * \frac{depth(lcs)}{(depth(X_j) + depth(Y_j))} \quad (5)$$

where lcs is the Least Common Subsumer of $X_j$ and $Y_j$, and depth is the number of nodes on the path from the node to the ontology root.

**Step 3: Reusing the Most Similar Case: Recommend Results**
In order to insure a better precision of the recommender results, the recommendation takes place only if the following condition is verified: $sim(S,S^p) \geq B$, where B is a threshold value. In this case, the system recommends the UP set of documents related to $S^p$ in the case base.

**Step 4: Revising the Proposed Solution**
The case base is updated based on the user feedback. If $sim(S,S^p) \neq j$, where j is the size of the vector S, then a new case is added to the case base. This case is composed of the current situation S and the user's preference UP containing the set of documents of the navigation made by the user in the current situation.

## 5. An illustrative scenario
The set of marketing people of a company can access to the most relevant data via their mobile phone. Paul is a sales representative of the company. Regarding Paul's agenda, he has a meeting with a wine client in Paris on Monday. When he arrives at his meeting, the system should recommend him the relevant information to better manage his meeting.
Table 1 describes the set C of cases $(S^i, UP^i)$, i=1, 2, 3, existing in the case base.

| S | | | | UP |
|---|---|---|---|---|
| situation | location | time | social | |
| $S^1$ | Evry | workday | manager | $UP^1$ |
| $S^2$ | Lille | holyday | friend | $UP^2$ |
| $S^3$ | Paris | workday | Champagne client | $UP^3$ |

**Table 1 The case base data**

When Paul arrives at his meeting, he uses his Smartphone to connect to his company data base and gets some information about the client.

The recommender system captures the current situation S (Paris, workday, Champagne client), as outlined in Step 1 of Section 3.2, and starts the CBR algorithm. Table 2 illustrates the ontologies used by the proposed CRS:

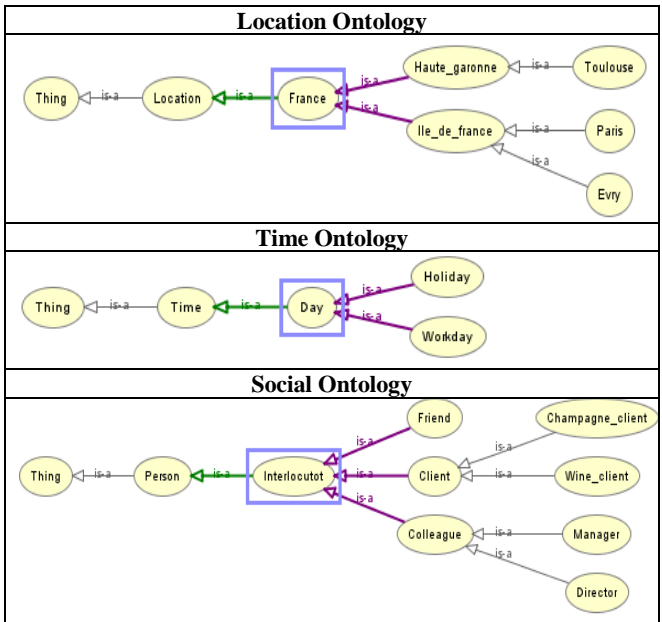

**Table 2 Location, Time and Social ontologies, resp.**

The algorithm computes the similarity between S and each of the situations in the case base $\{S^1, S^2, S^3\}$ by applying equation (5) (Step 2). As an exemple, let $S(X_1, X_2, X_3)$ and $S^1(Y_1, Y_2, Y_3)$, where $X_1$= « Paris », $Y_1$= « Evry », $Lcs$="Ile_de_france".

$$sim_1(X_1,Y_1) = 2*\frac{deph(Ile\_de\_france)}{(deph(Paris)+deph(Evry))} = 2*\frac{2}{(3+3)} = \frac{4}{6}$$

Doing the same for couples $(X_2, Y_2)$ and $(X_3, Y_3)$, we obtain $sim(S, S^1) = 4/6+1+2/6=2$. By computing the similarity for couples $(S, S^2)$ and $(S, S^3)$, we get vector V( 2, 1,23, 2,66). The CBR algorithm gets the argmax($V$), which is 2,66 corresponding to $S^3$. The next step is to reuse the most similar case by recommending the $UP^3$ to the user (Step 3).

Finally, the algorithm retains the case on the case base since $sim(S, S^3) \neq 3$ (Step 4).

## 5. Conclusions and Future Work

This paper describes a new approach for a CRS. It consists of three steps: (1) inferring semantic situations from low level social, location and time data, (2) learning and maintaining user's interests based on his navigation history related to the identified situations, and (3) selecting a preference to use for recommendation given a new situation by exploiting a CBR technique.

In the future, we plan to use our CRS in a real application in order to evaluate the impact of introducing the social-spatio-temporal user profiles in the recommender results.